# Low-dimensional modelling of a generalized Burgers equation

Zhenquan Li[1] and A.J. Roberts[2]

[1] Department of Mechanical Engineering, University of Auckland, Private Bag 92019, Auckland, New Zealand

[2] Department of Mathematics & Computing, University of Southern Queensland, Toowoomba 4350, Australia

*Abstract*

Burgers equation is one of the simplest nonlinear partial differential equations—it combines the basic processes of diffusion and nonlinear steepening. In some applications it is appropriate for the diffusion coefficient to be a time-dependent function. Using a Wayne's transformation and centre manifold theory, we derive l-mode and 2-mode centre manifold models of the generalised Burgers equations for bounded smooth time dependent coefficients. These modellings give some interesting extensions to existing results such as the similarity solutions using the similarity method.

## *1. INTRODUCTION*

Burgers equation

$$u_t + uu_x = cu_{xx} \quad (c \text{ is constant}) \qquad (1)$$

was originally introduced by J.M. Burgers (1939) as a model for turbulence [1]. It describes a variety of nonlinear wave phenomena arising in the theory of wave propagation, acoustics, plasma physics and other areas [30, 11]. It shares a number of properties with the Navier-Stokes equations: the same type of nonlinearity, of invariance groups and of energy dissipation relation, the existence of a multidimensional version, etc [12]. It is also one of the simplest nonlinear partial differential equations embodying together effects of convection and diffusion. For example, in an approximation to sound waves, the independent variable $x$ is a coordinate moving with the wave at the speed of sound, and the dependent variable $u$ represents velocity fluctuations [6].

M.J. Lighthil1 [15, p323] stated that the coefficient of the diffusion term $u_{xx}$ in equation (1) is not normally a constant in applications, even approximately; the coefficient may actually be a function of the time. There is significant interest in generalizing (1) to

$$u_t + uu_x = \frac{1}{2}\Delta(t)u_{xx} \qquad (2)$$

---

[1] Corresponding Author. Tel:+64-9373-7599 ext 4543. Fax: +64-9373-7479. E-mail: zhen.li@auckland.ac.nz

[2] E-mail: aroberts@usq.edu.au

for different functions $\Delta(t)$ [28, 27, 6, 26] , particularly in determining similarity solutions of (2). Such solutions are sought for compact initial conditions for *u,* but until our work here there has been a lack of assurance of the attractivity of the similarity solutions.

There is also a lot of work devoted to the decay of random solutions of (2) in the limit of vanishing $\Delta(t)$ as time tends to infinity [14, 10, 11, 18, 7, 8, 13, 12]. These works typically estimate the main quantities such as the energy spectrum, correlation function and so on for long times and the expressions are not dependent upon $\Delta(t)$. However, from the low-dimensional models in Section 3 most quantities should depend upon $\Delta(t)$ even if it vanishes for long times.

In this paper, we consider the diffusion coefficient
$$\Delta(t) = 2(\gamma + \delta t^r),$$
where *r* is non-positive real number, $\gamma$ is a non-negative real number and $\delta$ is a positive real number. Such a $\Delta(t)$ may be viewed as an approximation of a smoothly decaying diffusion.

We apply centre manifold theory to the derivation of the long-term solutions of (2). From the relevance theorem in [2], theory guarantees that any solution of (2) could be approximated to any order of error by the solution of the low-dimensional model on the centre manifold. Such centre manifolds are exponentially attractive so that we know a wide range of initial conditions approach the similarity solution.

The long-term solutions of the dynamics are derived through centre manifold theory after a variable transformation of the generalized Burgers equation. The main trick is the transformation $\tau = \log t$ which changes algebraic transients in *t* into exponential transients in $\tau$ and so allows centre manifold theory to be applied. A corresponding transformation of space is also needed. The low-dimensional models derived in Section 3 show the "amplitudes" *A* defined in Section 2 are near constant for long time. Therefore the dynamics of the generalized Burgers equation are determined by the centre manifolds in Section 3 for the diffusion coefficient $\Delta(t) = 2(\gamma + \delta t^r)$. In the case of $\gamma = 0$ and $r < 0$, the predictions depend upon how fast the diffusion coefficient vanish. However, the predictions in other papers (for example, [12]) show results for the general case of diffusion coefficient $\Delta(t) = \varepsilon$ where $\varepsilon \to 0$ as $t \to 0$.

## *2. CENTER MANIFOLD APPROXIMATION*

In this section, we first introduce a transformation of the generalized Burgers equation (2). We then discuss the properties of its solutions using center manifold theory.

Define new variable through Wayne's transformation [29] $\tau = \log t$, $\xi = xt^{-\beta}$, and $u = Ct^{-\alpha}v(\xi, \tau)$ for some $C > 0$, $\alpha > 0$, and $\beta > 0$. Algebraic transients in *t* become exponential transients in $\tau = \log t$ allowing center manifold theory to be applied; $\xi = xt^{-\beta}$ stretches space so that a spreading Gaussian will become a fixed point of the dynamics. The investigation of the dynamics is based on this Gaussian. M.J. Lighthill [15, p335] showed that the form of solutions of Burgers equation with constant diffusion coefficient are roughly Gaussian. The generalized Burgers equation (2) under the above transformation becomes

$$\frac{\partial v}{\partial \tau} - \alpha v - \beta \xi \frac{\partial v}{\partial \xi} + C t^{-\alpha - \beta + 1} v \frac{\partial v}{\partial \xi} = \frac{1}{2} \Delta(t) t^{-2\beta + 1} \frac{\partial^2 v}{\partial \xi^2}. \tag{3}$$

We consider the diffusion coefficient in the form
$$\Delta(t) = 2(\gamma + \delta t^r) \tag{4}$$
where $r$ is a non-positive real number, $\gamma$ is a non-negative real number and $\delta$ is a positive real number. This may be viewed as an approximation to wider class of functions $\Delta(t)$. We lay the basis for two different models for the large-time dynamics depending upon whether $\gamma = 0$ or not, that is, depending upon whether or not the diffusion coefficient $\Delta \to 0$ for large time.

A. If $\gamma \neq 0$, take $-\alpha - \beta + 1 = 0$ and $-2\beta + 1 = 0$, i.e., $\alpha = \beta = 1/2$, then (3) becomes
$$\frac{\partial v}{\partial \tau} = \frac{1}{2} v + \frac{1}{2} \xi \frac{\partial v}{\partial \xi} + \gamma \frac{\partial^2 v}{\partial \xi^2} - C v \frac{\partial v}{\partial \xi} + \delta t^r \frac{\partial^2 v}{\partial \xi^2}. \tag{5}$$

Introducing $\theta = \frac{\delta}{2\gamma} t^r$, we obtain the equivalent system:
$$\frac{\partial v}{\partial \tau} = \left( \gamma \frac{\partial^2}{\partial \xi^2} + \frac{1}{2} \xi \frac{\partial}{\partial \xi} + \frac{1}{2} \right) v - C v \frac{\partial v}{\partial \xi} + 2\gamma \theta \frac{\partial^2 v}{\partial \xi^2}, \qquad \frac{\partial \theta}{\partial \tau} = r\theta. \tag{6}$$

Rescaling $\varsigma = \sqrt{\frac{1}{2\gamma}} \xi$, $\tau' = \tau/2$ and choosing $C = \sqrt{\frac{\gamma}{2}}$ the system (6) becomes
$$\frac{\partial v}{\partial \tau'} = \Im_1 v - v \frac{\partial v}{\partial \varsigma} + 2\theta \frac{\partial^2 v}{\partial \varsigma^2}, \qquad \frac{\partial \theta}{\partial \tau'} = 2r\theta, \tag{7}$$

where the operator
$$\Im_\sigma = \frac{\partial^2}{\partial \varsigma^2} + \varsigma \frac{\partial}{\partial \varsigma} + \sigma,$$
with $\sigma = 1$ for this case.

B. If $\gamma = 0$, take $r - 2\beta + 1 = 0$, i.e., $\beta = (1 + r)/2$ and let $\theta = \sqrt{\frac{\beta}{\delta}} t^p$, where $p = -\alpha - \beta + 1 = 1/2 - \alpha - r/2$, and $\alpha$ will be decided later, then (3) becomes
$$\frac{\partial v}{\partial \tau} = \left( \delta \frac{\partial^2}{\partial \xi^2} + \beta \xi \frac{\partial}{\partial \xi} + \alpha \right) v - C \theta v \frac{\partial v}{\partial \xi}, \qquad \frac{\partial \theta}{\partial \tau} = p\theta. \tag{8}$$

Rescaling $\varsigma = \sqrt{\frac{\beta}{\delta}} \xi$, $\tau' = \beta \tau$ and choosing $C = \sqrt{\delta \beta}$ the system (8) becomes
$$\frac{\partial v}{\partial \tau'} = \Im_\sigma v - \theta v \frac{\partial v}{\partial \varsigma}, \qquad \frac{\partial \theta}{\partial \tau'} = \frac{r}{\beta} \theta, \tag{9}$$
where $\sigma = \alpha/\beta = 2\alpha/(1 + r)$.

Both systems (7) and (9) have the same from for the linear parts. So we discuss systems in the general form
$$\frac{\partial w}{\partial \tau'} = \Im_\sigma w + f(w, \theta), \qquad \frac{\partial \theta}{\partial \tau'} = c\theta, \tag{10}$$
where $f(w, \theta)$ are some nonlinear terms, and $c$ is a constant.

The application of center manifold theory hinges upon the spectrum of the linearised dynamics. Seek the solution of the linear pat of the first equation in system (10) in the form $w = \exp(\lambda\tau' - \varsigma^2/2)\phi(\varsigma)$. Substituting into the linear part of (10), we obtain a Hermite differential equation for $\phi(\varsigma)$

$$\frac{d^2\phi}{d\varsigma^2} - \varsigma\frac{d\phi}{d\varsigma} + l\phi = 0, \qquad (11)$$

where $l = \sigma - 1 - \lambda$. The sequence of well-behaved solutions of (11) are Hermite polynomials $H_l(\varsigma)$ for $l = 0, 1, 2, \cdots$. Thus the spectrum of $\Im_\sigma$ consists of eigenvalues $\lambda_l = \sigma - 1 - l$, $l = 0, 1, 2, \cdots$, with corresponding eigenvectors $H_l(\varsigma)\exp(\lambda_l\tau' - \varsigma^2/2)$. Therefore there exists a center manifold for system (10) iff $\sigma = 1$ by the existence theorem of center manifold theory. Linearly, all modes decay exponentially in $\tau'$ except for the critical mode, $H_0(\varsigma)\exp(-\varsigma^2/2)$, which is constant in time.

Use these general results to deduce specific properties of systems (7) and (9).

- For the system (7), the spectrum of $\Im_1$ consists of $\lambda_l = -l$, $l = 0, 1, 2, \cdots$. Therefore all modes decay exponentially quickly except fir the critical mode $H_0(\varsigma)\exp(-\varsigma^2/2)$, it has zero decay rate and thus is long lasting. Note $H_0$ is constant. We have the following three cases to discuss for system (7):
  1) There exists one critical mode for $r < 0$, namely $v = A\exp(-\varsigma^2/2)$, $\theta = 0$;
  2) There exist two critical modes for $r \approx 0$, namely $v = A\exp(-\varsigma^2/2)$, and $\theta$;
  3) And when $r > 0$ there exists a two dimensional center-unstable manifold based upon the same two modes. This case will be disused later as here we assume $r$ is non-positive (i.e. decreasing diffusity $\Delta(t)$).

- For the system (9), the spectrum of $\Im_1$ consists of $\lambda_l = -l$, $l = 0, 1, 2, \cdots$ upon choosing $\alpha = (1+r)/2$, for any given $r > -1$. The restriction of $r > -1$ comes from the first case of the following three cases. Then only the critical mode $H_0(\varsigma)\exp(-\varsigma^2/2)$ is long lasting. We discuss the following three cases for system (9):
  1) There exists one critical mode for $-1 < r < 0$, since $\tau' = \tau(1+r)/2$ and $\tau$ must keep the same sign as $\tau'$;
  2) There exist two critical modes for $r \approx 0$;
  3) There exists a two dimensional center-unstable manifold for $r > 0$. This case will be discussed later as here we assume $r$ is non-positive (i.e. decreasing diffusity $\Delta(t)$).

Now we return our attention to the general system (10) and discuss the first two cases of both systems (7) and (9), i.e. we consider the following two cases:
1) There exists one critical mode for $c < 0$;
2) There exist two critical modes for $c \approx 0$.

Due to the special form of the solution, we model the dynamics in terms of the evolution of "amplitude" $A$ of the Gaussian $\exp(-\varsigma^2)$, where we define $A = \frac{1}{\sqrt{\pi}} \int_{-\infty}^{\infty} v \exp(-\varsigma^2/2) d\varsigma$. The definition of $A$ is different from the usual one of amplitude. To improve the convergence of the improper integral used in this paper, we multiply $v$ by $\exp(-\varsigma^2)$ in the definition. We call the "$\exp(-\varsigma^2)$" the weight of the integral. Note that $A$ is only the function of $\tau'$ for the first case since $\theta = 0$ for this case. $A$ is a function of $\tau'$ and $\theta$ for the second case. The evolution of $A(\tau')$ and $A(\tau', \theta)$ form accurate one-dimensional models of the infinite-dimensional dynamical system (10).

## 3. LOW-DIMENSIONAL MODEL OF THE GENERALIZED BURGERS EQUATION

The derivation of the low-dimensional model of the system (10) for the first two cases is done with an iteration scheme. The predictions of these models are discussed following the derivation of the models. Roberts [23] developed such an iteration scheme using computer algebra. We summarize the procedure for system (10) as follows. We here give the procedure for Case 2. There are two critical modes $v = A\exp(-\varsigma^2/2)$ and $\theta$ in the second case, both of which are included in the iteration scheme. Therefore, the iteration scheme for the first case is a special situation of the one for the second case simply by omitting the expressions for $\theta$.

1) Identify the critical mode or modes, i.e., find the nontrivial solutions $w$, of $\Im_1 w = 0$, to give the linear approximation

$$w \approx A(\tau', \theta)\exp(-\varsigma^2/2), \qquad \text{such that } \dot{A} \approx \dot{\theta} \approx 0, \qquad (12)$$

where $\dot{A}$ denotes $\frac{\partial A}{\partial \tau'}$.

2) Find a low-dimensional description which satisfies the system (10). The details of this step are described as follows. Suppose that we have some approximate model

$$w \approx \tilde{w}(A, \theta), \quad \text{such that } \dot{A} \approx \tilde{h}(A, \theta) \text{ and } \dot{\theta} = c\theta \qquad (13)$$

at any stage of the iteration scheme. Then we seek corrections so that a better approximation to the center manifold and the evolution is

$$w = \tilde{w}(A, \theta) + w'(A, \theta), \qquad \text{such that } \dot{A} = \tilde{h}(A, \theta) + h'(A, \theta).$$

For example, in the first iteration,

$$w = A\exp(-\varsigma^2/2) + w'(A, \theta), \qquad \text{and} \qquad \dot{A} \approx h'(A, \theta).$$

Substituting into system (10), and using the chain rule for time derivatives leads to

$$\left(\frac{\partial \tilde{w}}{\partial A} + \frac{\partial w'}{\partial A}\right)(\tilde{h} + h') + \left(\frac{\partial \tilde{w}}{\partial \theta} + \frac{\partial w'}{\partial \theta}\right)c\theta = \Im_1(\tilde{w} + w') + f(\tilde{w} + w', \theta).$$

The difficulty of solving the above equation is almost the same as that of solving the system (10), so further simplification is appropriate. Ignoring the products of corrections because they are small compared with other terms, and using the linear approximation wherever terms multiply corrections, we obtain an equation for the correction

$$\mathfrak{I}_1 w' + \exp(-\varsigma^2/2) h' = \tilde{r}, \tag{14}$$

where $\tilde{r}$ is the residual of the first equation in (10). Equation (14) is solved by the following scheme for the corrections: first choose $h'$ to put $\tilde{r} - \exp(-\varsigma^2/2) h'$ in the range of $\mathfrak{I}_1$; and second solve $\mathfrak{I}_1 w' = rhs$. The iteration is carried out until the residual $\tilde{r}$ becomes zero to a chosen order of error. Note that the condition $A = \dfrac{1}{\sqrt{\pi}} \int_{-\infty}^{\infty} v \exp(-\varsigma^2/2) d\varsigma$ must be satisfied.

A REDUCE program given in the appendix A is run to perform the computations of the low dimensional model for the Case 1 of system (7). In this program, $\mathfrak{I}_\sigma^{-1}(\varsigma^l \exp(k\varsigma^2/2))$ $(l=1,2,\cdots, k=1,2,\cdots,5)$ is expanded by products of $\exp(-k\varsigma^2/2)$ $(k=1,2,\cdots,5)$ and a series in $\varsigma$ with error $\mathbf{O}(\xi^8)$[3]. The REDUCE programs for other cases are obtained by a few changes of the program given in the Appendix A. From the further calculations by the program in Appendix A, we know that all the radii of coefficient series in $\varsigma$ of $\exp(-k\varsigma^2/2)$ $(k=1,2,\cdots,5)$ are $+\infty$.

## 3.1 Case $\gamma \neq 0$

In this subsection, we represent the center manifolds and the corresponding low-dimensional models for $\gamma \neq 0$.

The center manifold for $r < 0$ is

$$\begin{aligned} v = {} & GA - G^2 A^2 \left(0.1667\varsigma^3 + 0.0833\varsigma^5 + 0.0238\varsigma^7\right) \\ & - GA^3 \left(0.0090 - 0.0321\varsigma^2 - 0.0053\varsigma^4 - 0.0007\varsigma^6\right) \\ & - G^3 A^3 \left(0.0417\varsigma^4 + 0.0278\varsigma^6\right) + G^2 A^4 \left(0.0137\varsigma^3 + 0.0036\varsigma^5 + 0.001\varsigma^7\right) \\ & - G^4 A^4 \left(0.0083\varsigma^5 + 0.0056\varsigma^7\right) + GA^5 \left(0.004 - 0.0020\varsigma^2 + 0.0002\varsigma^4 + 0.0001\varsigma^6\right) \\ & + G^3 A^5 \left(0.0038\varsigma^4 + 0.0008\varsigma^6\right) - 0.0014 G^5 A^5 \varsigma^6 + \mathbf{O}\left(\varsigma^8, A^6\right) \end{aligned}$$
(15)

on which the system evolves according to

$$\frac{dA}{d\tau'} = 0.0641 A^3 - 0.0022 A^5 + \mathbf{O}(A^6), \tag{16}$$

where $G = \exp(-\varsigma^2/2)$. Then the center manifold of the Burgers equation with the diffusion coefficient (4) for $r < 0$ is obtained by substituting $\varsigma = \sqrt{\dfrac{1}{2\gamma}} x t^{-1/2}$ into equation (15), and then substituting equation (15) into $u = \sqrt{\dfrac{\gamma}{2}} t^{-1/2} v$. The error of the center manifold is $\mathbf{O}(x^8 t^{-4}, A^6)$. On this center manifold the system evolves according to

$$\frac{dA}{dt} = 0.0321 A^3 t^{-1} - 0.0011 A^5 t^{-1} + \mathbf{O}(A^6 t^{-1}). \tag{17}$$

---

[3] A proof of the uniform convergence of such series of functions in $(-\infty, +\infty)$ and integral by term is given in Appendix B.

The state of $A$ described by equation (17) on the center manifold is approached with transients of relative magnitude approximately $\exp(-\tau') \propto t^{-1/2}$. The dynamics of the Burgers equation with the diffusion (4) for $\gamma \neq 0$ and $r < 0$ evolves for long time according to the expression of $u$ in $x, t$ and $A$ ($A$ is the solution of (17)).

We consider $\theta$ as a perturbation for the system when $r \approx 0$. The center manifold for the case 2 of system (7) is

$$v = GA\left[1 - \theta\left(0.1257 - 0.4082\varsigma^2 + 0.0986\varsigma^4 + 0.0132\varsigma^6\right)\right]$$
$$+ GA^3\left[-0.0090 + 0.0321\varsigma^2 + 0.0053\varsigma^4 + 0.0007\varsigma^6\right.$$
$$\left. + \theta\left(0.0361 - 0.1427\varsigma^2 - 0.0136\varsigma^4 - 0.0030\varsigma^6\right)\right]$$
$$+ GA^5\left[0.0004 - 0.0020\varsigma^2 + 0.0002\varsigma^4 + 0.0001\varsigma^6\right.$$
$$\left. - \theta\left(0.0029 - 0.0156\varsigma^2 + 0.0022\varsigma^4 + 0.0004\varsigma^6\right)\right]$$
$$- G^2A^2\left[0.1667\varsigma^3 + 0.0833\varsigma^5 + 0.0238\varsigma^7\right.$$
$$\left. - \theta\left(0.5113\varsigma^3 + 0.1482\varsigma^5 + 0.0317\varsigma^7\right)\right]$$
$$+ G^2A^4\left[0.0137\varsigma^3 + 0.0036\varsigma^5 + 0.0003\varsigma^7\right.$$
$$\left. - \theta\left(0.0899\varsigma^3 + 0.0153\varsigma^5 + 0.0009\varsigma^7\right)\right]$$
$$- G^3A^3\left[0.0417\varsigma^4 + 0.0278\varsigma^6 - \theta\left(0.2164\varsigma^4 + 0.1061\varsigma^6\right)\right]$$
$$+ G^3A^5\left[0.0038\varsigma^4 + 0.0008\varsigma^6 - \theta\left(0.0332\varsigma^4 + 0.0033\varsigma^6\right)\right]$$
$$- G^4A^4\left[0.0083\varsigma^5 + 0.0056\varsigma^7 - \theta\left(0.0297\varsigma^7\right)\right]$$
$$- G^5A^5\left[0.0014\varsigma^4 - \theta\left(0.0131\varsigma^6\right)\right] + \mathbf{O}\left(\varsigma^8, A^6, \theta^2\right), \tag{18}$$

on which the system evolves according to

$$\frac{dA}{d\tau'} = -1.1835A\theta + A^3(0.0641 - 0.1631\theta) - A^5(0.0022 - 0.0133\theta) + \mathbf{O}\left(A^6, \theta^2\right). \tag{19}$$

Then the center manifold of the Burgers equation with the diffusion coefficient (4) for $r \approx 0$ is obtained by substituting $\varsigma = \sqrt{\frac{1}{2\gamma}} x t^{-1/2}$ into equation (18), and then substituting equation (18) into $u = \sqrt{\frac{\gamma}{2}} t^{-1/2} v$. The error of the center manifold is $\mathbf{O}\left(x^8 t^{-4}, A^6, \theta^2\right)$. On this center manifold the system evolves according to

$$\frac{dA}{dt} = -0.5918A\theta t^{-1} + A(0.0321 - 0.0816\theta)t^{-1} - A^5(0.0011 - 0.0067\theta)t^{-1}$$
$$+ \mathbf{O}(A^6 t^{-1}, \theta^2 t^{-1}).$$
(20)

The state of $A$ described by (20) on the center manifold is approached with transients of relative magnitude approximately $\exp(-\tau') \propto t^{-1/2}$. The dynamics of the Burgers equation with diffusion coefficient (4) for $\gamma \neq 0$ and $r \approx 0$ evolves for long time according to the expression of $u$ in $x$, $\theta$, $t$ and $A$ ($A$ is the solution of (20)).

## 3.2 Case $\gamma = 0$

In this subsection, we give the center manifold and the corresponding low-dimensional model for $\gamma = 0$.

The center manifold for the case 1 of system (9) is
$$v = AG,$$
(21)
on which the system evolves according to
$$\frac{dA}{d\tau'} = 0.$$
(22)

Then the center manifold of the Burgers equation with the diffusion coefficient (4) for $r < 0$ is obtained by substituting $\varsigma = \sqrt{\frac{1+r}{2\delta}} x t^{-(1+r)/2}$ into equation (21), and then substituting equation (21) into $u = \sqrt{\frac{\delta(1+r)}{2}} t^{-(1+r)/2} v$, i.e.,

$$u = A\sqrt{\frac{\delta(1+r)}{2}} t^{-(1+r)/2} \exp\left(-\frac{(1+r)x^2}{4\delta t^{1+r}}\right),$$
(23)

on which the system evolves according to
$$\frac{dA}{dt} = 0.$$
(24)

Equation (24) indicates that the amplitude $A$ in (23) is a constant. The state of $A$ on the center manifold is approached with transients of relative magnitude approximately $\exp(-\tau') \propto t^{-(1+r)/2}$. The dynamics of the Burgers equation with the diffusion coefficient (4) for $\gamma = 0$ and $r < 0$ evolves for long time according to (23).

As in subsection 3.1, we still consider $\theta$ as a perturbation for the system when $r \approx 0$. The center manifold for the case 2 of system (9) is

$$v = GA - GA^3\theta^2\left(0.0090 - 0.0321\varsigma^2 - 0.0053\varsigma^4 - 0.0007\varsigma^6\right)$$
$$- G^2 A^2\theta\left(0.1667\varsigma^3 + 0.0833\varsigma^5 + 0.0238\varsigma^7\right)$$
$$- G^3 A^3\theta^2\left(0.0417\varsigma^4 + 0.0278\varsigma^6\right)$$
$$+ \mathbf{O}\left(\varsigma^8, A^6, \theta^3\right)$$
(25)

on which the system evolves according to
$$\frac{dA}{d\tau'} = 0.0641 A^3 \theta^2 + \mathbf{O}\left(A^6, \theta^3\right).$$
(26)

Then the center manifold pf the Burgers equation with the diffusion coefficient (4) for $r \approx 0$ is obtained by substituting $\varsigma = \sqrt{\dfrac{1+r}{2\delta}} x t^{-(1+r)/2}$ into equation (25), and then substituting equation (25) into $u = \sqrt{\dfrac{\delta(1+r)}{2}} t^{-(1+r)/2} v$. The error of the center manifold is $\mathbf{O}\left(x^8 t^{-4(1+r)}, A^6, \theta^3\right)$. On this center manifold the system evolves according to

$$\frac{dA}{dt} = 0.0321(1+r)A^3\theta^2 t^{-1} + \mathbf{O}\left(A^6 t^{-1}, \theta^3 t^{-1}\right). \tag{27}$$

The state of $A$ described by equation (27) on the center manifold is approached with transients of relative magnitude approximately $\exp(-\tau') \propto t^{-(1+r)/2}$. The dynamics of the Burgers equation with the diffusion coefficient (4) for $\gamma = 0$ and $r \approx 0$ evolves for long time according to the expression of $u$ in $x$, $\theta$, $t$ and $A$ ($A$ is the solution of (27)).

## *APPENDIX A: COMPUTER ALGEBRA IMPLEMENTAION*

```
1 Comment. Find low-dimensional model for generalized Burgers equation
2  AM(t) measures amplitude of exp(-\xi-2/2) component in v(\xi,\tau)
3  x=\zeta, t=\rho.
4  Solve approximately in power series in x times powers of
5  exp(-x^2/2) for the case 1 of system (7).
6  on rounded;
7  on div; off allfac; on revpri; %for improving appearance of output
8  factor ga,am,theta;
9  0:=8;
10 procedure ignore-order-x(o);
11 begin
12 IF 0=6 THEN LET x^6=0;
13 IF 0=7 THEN LET x^7=0;
14 IF 0=8 THEN LET x^8=0;
15 IF 0=9 THEN LET x^9=0;
16 IF 0=10 THEN LET x^10=0;
17 end;
18 % define ga with properties of exp(-k*x^2/2)operator ga;
19 operator ga; depend ga,x
20 let{ df(ga(~k),x) => -k*x*ga(k),
21      ga(~k)^2 => ga(2*k),
22      ga(~k)*ga(~l) => ga(k+l)
23     };
24 %Define the inverse operator of {\cal J}-\sigma mod \zeta^8 in(7)
25 operator linv; linear linv;
26 let{linv(x^~a*ga(~b),x)=>(x^(a+2)*ga(b)-linv(b*(b-1)*x^(a+4)*ga(b)
27             +(3+a-5*b-2*a*h)*x^(a+2)*ga(b),x))/(a+2)/(a+1)
28             when evenp(a)or(b>1),
29     linv(x*ga(~b),x) => (x^3*ga(b)-linv(b*(b-1)*x^5*ga(b)
30             +(4-7*b)*x^3*ga(b),x))/6 when b>1,
31     linv(ga(~b),x) => (x^2*ga(b)-linv(b*(b-1)*x^4*ga(b)
32             +(3-5*b)*x-2*ga(b),x))/2,
33   linv(x~^a*ga(1),x)=>(x^a*ga(1)-linv(x^(a-2)*ga(1)*(a^2-
               a),x))/(-a)
34                when not evenp(a),
35   linv(x*ga(1),x) => -x*ga(1)
36     };
37 % Define integral for ga(k) from -\infi to \infi to get h
38 operator intg; linear intg;
39 let {intg(ga(~k),x) => sqrt(2*pi/(k+1)),
40      intg(ga(~k)*x,x) => 0,
41      intg(ga(~k)*x~^p,x) => (p-1)/(k+1)*intg(ga(k+1)*x^(p-2),x)
42     };
43 depend am,t;   % asserts that A depend on pseudotime \tau
44 let df(am,t) => h;   % dAM/d\tau is replaced by function h(AM)
45 depend theta,t;
46 let df(theta,t)=>2*r*theta;
47 v:=AM*ga(1); theta:=O; h:=O; % initial approximation
48 %
49 % iterate until PDE is satisfied to desired precision
50 let {AM^6=0,theta~2=0}; % discard high-order terms in AM
51 repeat begin
52    eqn:=df(v,t)-df(v,x,x)-x*df(v,x)-v+v*df(v,x)
53            -2*theta*df(v,x,x);
54    ignore-order-x(o);
55    eqn:=eqn;
56    eqt:=df(theta,t)-2*r*theta;
```

```
57      ignore-order-x(o+3);
58      gh:=-1/sqrt(pi)*intg(eqn,x);
59      vd:=linv(eqn+gh*ga(1),x);
60      v:=v+vd-intg(vd,x)*ga(1)/sqrt(pi);
61      h:=h+gh;
62   end until eqn=0 and eqt=0;
63   % check amplitude
64   amp:=intg(v,x)-sqrt(pi)*AM;
65   ;end;
```

## *APPENDIX B: PROOF OF THE ASYMPTOTIC SCHEME FOR BURGERS EQUATION*

In Section 3, we need to calculate $\Im_\sigma^{-1}\left(\varsigma^l \exp\left(-k\varsigma^2/2\right)\right)$ to get the center manifold. However, it is hard to do that. We use a series of function to approximate it, i.e. let $\Im_\sigma^{-1}\left(\varsigma^l \exp\left(-k\varsigma^2/2\right)\right) = f_{lk}(\varsigma)$, we seek a series of function:

$$f_{lk}(\varsigma) = f_{lk}^1(\varsigma) + f_{lk}^2(\varsigma) + \cdots + f_{lk}^n(\varsigma) + \cdots$$
$$= G^k\left(a_0 + a_1\varsigma + \cdots + a_n\varsigma^n + \cdots\right) \tag{28}$$

where $G = \exp\left(-\varsigma^2/2\right)$. In this section, we first prove that the series of function in the right hand side of (28) is uniformly convergent to $f_{lk}(\varsigma)$ in $(-\infty, +\infty)$. Thus we can replace $f_{lk}(\varsigma)$ by finite terms of the series in the right hand side of (28) in any order of accuracy. We also use the integral of $f_{lk}(\varsigma)G$ from $-\infty$ to $+\infty$ in the calculation of the center manifold and the low-dimensional models. Then we prove that the numerical series of the integral of the product of every term in the right hand side of (28) with $G$ from $-\infty$ to $+\infty$ is convergent and converges to the integral of $f_{lk}(\varsigma)G$ from $-\infty$ to $+\infty$. Note that $G$ is the weight of the integral.

Without loss of generality, we prove the results for $G$. The proofs of the others are the same.

It is easy to calculate:

$$\Im_\sigma^{-1}(G) = Gf_1(\varsigma)$$
$$= G\left(\frac{1}{2}\varsigma^2 + \frac{1}{12}\varsigma^4 + \cdots + \frac{1}{c_{2n}}\varsigma^{2n} + \cdots\right)$$

where $c_{2n} = 2n(2n-1)c_{2(n-1)}/2(n-1)$ for $n > 1$. From the Ratio test of series of functions,

$$\lim_{n\to\infty}\left|\frac{u_n}{u_{n-1}}\right| = \lim_{n\to\infty}\frac{\varsigma^2 2(n-1)}{2n(2n-1)} = 0,$$

series $f_1(\varsigma)$ converges for all $\varsigma$, i.e. series $Gf_1(\varsigma)$ uniformly converge to $\Im_\sigma^{-1}(G)$ in $(-\infty, +\infty)$.

Now let us turn our attention to prove

$$\int_{-\infty}^{+\infty} \mathfrak{I}_\sigma^{-1}(G)G d\varsigma = \int_{-\infty}^{+\infty} G^2 f_1(\varsigma) d\varsigma$$

$$= \frac{1}{2}\int_{-\infty}^{+\infty}\varsigma^2 G^2 d\varsigma + \frac{1}{12}\int_{-\infty}^{+\infty}\varsigma^4 G^2 d\varsigma + \cdots + \frac{1}{c_{2n}}\int_{-\infty}^{+\infty}\varsigma^{2n} G^2 d\varsigma + \cdots.$$

(29)

According to the definition of improper integral in $(-\infty, +\infty)$,

$$\int_{-\infty}^{+\infty} f(x) dx = \int_{-\infty}^{0} f(x) dx + \int_{0}^{+\infty} f(x) dx$$

if the two improper integrals on the right hand side of above equation are convergent. However, the function $f$ we considered in this paper is even or odd. Then the first integral on the right hand side can be transformed to the product of a constant and the second one. Thus it is enough to give the proof of (29) in $[0, +\infty)$. For $n > 0$,

$$\int_0^{+\infty} \varsigma^{2n} G^2 d\varsigma = \int_0^{+\infty} \varsigma^{2n} \exp(-\varsigma^2) d\varsigma$$

$$= -2^{-1}\varsigma^{2n-1}\exp(-\varsigma^2)\Big|_0^{+\infty} + 2^{-1}(2n-1)\int_0^{+\infty}\varsigma^{2(n-1)}\exp(-\varsigma^2)d\varsigma$$

$$= \cdots$$

$$= 2^{-n}(2n-1)(2n-3)\cdots 1 \int_0^{+\infty} \exp(-\varsigma^2) d\varsigma$$

$$= 2^{-n}(2n-1)(2n-3)\cdots 1 \frac{\sqrt{\pi}}{2}.$$

Apply the Ratio test of series of constants to the right hand side of (29),

$$\lim_{n\to\infty}\left|\frac{u_n}{u_{n-1}}\right| = \lim_{n\to\infty}\frac{n-1}{2n} = \frac{1}{2},$$

the series of the right hand side of (29) is convergent. Let the sum of the series is $S$. Thus for $\forall \varepsilon > 0$, $\exists$ integer $N$, when $n > N$,

$$\left|\frac{1}{c_{2n}}\int_0^{+\infty}\varsigma^{2n}G^2 d\varsigma + \frac{1}{c_{2(n+1)}}\int_0^{+\infty}\varsigma^{2(n+1)}G^2 d\varsigma + \cdots\right| < \varepsilon/2.$$

So

$$\left|\frac{1}{c_{2(N+1)}}\int_0^{+\infty}\varsigma^{2(N+1)}G^2 d\varsigma + \frac{1}{c_{2(N+2)}}\int_0^{+\infty}\varsigma^{2(N+2)}G^2 d\varsigma + \cdots\right| < \varepsilon/2. \qquad (30)$$

Since $\frac{1}{c_2}\varsigma^2 G^2, \cdots, \frac{1}{c_{2N}}\varsigma^{2N} G^2$ are integrable in $[0, +\infty)$, then $\exists M_0 > 0$, when $M > M_0$,

$$\left|\int_M^{+\infty}\frac{1}{c_{2i}}\varsigma^{2i} G^2 d\varsigma\right| < \frac{\varepsilon}{2N} \qquad (i = 1, 2, \cdots, N). \qquad (31)$$

Since the series of (28) is uniform convergent and each term of the series is continuous and positive in $[0, +\infty)$, then

$$\int_0^M \mathfrak{I}_\sigma^{-1}(G)Gd\varsigma = \frac{1}{2}\int_0^M \varsigma^2 G^2 d\varsigma + \frac{1}{12}\int_0^M \varsigma^4 G^2 d\varsigma + \cdots + \frac{1}{c_{2n}}\int_0^M \varsigma^{2n} G^2 d\varsigma + \cdots, \quad (32)$$

and

$$\left|\int_0^M \mathfrak{I}_\sigma^{-1}(G)Gd\varsigma\right| \leq \left|\frac{1}{2}\int_0^M \varsigma^2 G^2 d\varsigma\right| + \left|\frac{1}{12}\int_0^M \varsigma^4 G^2 d\varsigma\right| + \cdots + \left|\frac{1}{c_{2n}}\int_0^M \varsigma^{2n} G^2 d\varsigma\right| + \cdots$$

$$\leq \left|\frac{1}{2}\int_0^{+\infty} \varsigma^2 G^2 d\varsigma\right| + \left|\frac{1}{12}\int_0^{+\infty} \varsigma^4 G^2 d\varsigma\right| + \cdots + \left|\frac{1}{c_{2n}}\int_0^{+\infty} \varsigma^{2n} G^2 d\varsigma\right| + \cdots$$

$$= \frac{1}{2}\int_0^{+\infty} \varsigma^2 G^2 d\varsigma + \frac{1}{12}\int_0^{+\infty} \varsigma^4 G^2 d\varsigma + \cdots + \frac{1}{c_{2n}}\int_0^{+\infty} \varsigma^{2n} G^2 d\varsigma + \cdots$$

$$= S.$$

Also since

$$\int_0^M \mathfrak{I}_\sigma^{-1}(G)Gd\varsigma = \left|\int_0^M \mathfrak{I}_\sigma^{-1}(G)Gd\varsigma\right|$$

increases when $M$ increases, then

$$\int_0^{+\infty} \mathfrak{I}_\sigma^{-1}(G)Gd\varsigma \text{ exists and } \leq S. \quad (33)$$

On the other hand, from (32) and the conditions

$$\int_0^M \mathfrak{I}_\sigma^{-1}(G)Gd\varsigma = \frac{1}{2}\int_0^M \varsigma^2 G^2 d\varsigma + \frac{1}{12}\int_0^M \varsigma^4 G^2 d\varsigma + \cdots + \frac{1}{c_{2n}}\int_0^M \varsigma^{2n} G^2 d\varsigma + \cdots$$

$$= \frac{1}{2}\int_0^{+\infty} \varsigma^2 G^2 d\varsigma + \frac{1}{12}\int_0^{+\infty} \varsigma^4 G^2 d\varsigma + \cdots + \frac{1}{c_{2n}}\int_0^{+\infty} \varsigma^{2n} G^2 d\varsigma + \cdots$$

$$- \left[\frac{1}{2}\int_M^{+\infty} \varsigma^2 G^2 d\varsigma + \frac{1}{12}\int_M^{+\infty} \varsigma^4 G^2 d\varsigma + \cdots + \frac{1}{c_{2n}}\int_M^{+\infty} \varsigma^{2n} G^2 d\varsigma + \cdots\right]$$

$$\geq S - \left[\frac{1}{2}\int_M^{+\infty} \varsigma^2 G^2 d\varsigma + \frac{1}{12}\int_M^{+\infty} \varsigma^4 G^2 d\varsigma + \cdots + \frac{1}{c_{2N}}\int_M^{+\infty} \varsigma^{2N} G^2 d\varsigma\right]$$

$$- \left[\frac{1}{c_{2(N+1)}}\int_M^{+\infty} \varsigma^{2(N+1)} G^2 d\varsigma + \cdots\right].$$

When $M > M_0$, the part in the bracket of the second term in above expression is less than $\varepsilon/2$ from (31). The part in the bracket of the third term is less than $\varepsilon/2$ from (30). Thus

$$\int_0^{+\infty} \mathfrak{I}_\sigma^{-1}(G)Gd\varsigma \geq S - \varepsilon.$$

Because ε is arbitrary positive, we have

$$\int_0^{+\infty} \Im_\sigma^{-1}(G) G d\varsigma \geq S. \qquad (34)$$

Combining (33) and (34), (29) holds.

## *References*

[1] J. M. Burgers, Mathematical examples illustrating relations occurring in the theory of turbulent fluid motion, *Kon. Ned. Akad. Wet. Verh* 17 (1939) 1—53.

[2] J. Carr, *Applications of centre manifolds theory,* Appl. Math. Sci., Vol. 35 (Springer, Berlin, 1981).

[3] H.C. Chang, Wave evolution on a falling film, *Annu. Rev. Fluid Mech.* 26 (1994) 103—136.

[4] S.M. Cox, A.J. Roberts, Centre manifolds of forced dynamical systems, *J. Austral. Math. Soc.* B 32 (1991) 401—436.

[5] S.M. Cox, A.J. Roberts, Initial conditions for modes of dynamical systems, *Physica* D, 85 (1995) 126—141.

[6] J. Doyle, M. J. Englefield, Similarity solutions of a generalized Burgers equation, *IMA Journal of Applied Mathematics* 44 (1990) 145—153.

[7] S.E. Esipov, T.J. Newman, Interface growth and Burgers turbulence: the problem of random initial conditions, *Phys. Rev.* E 48 (1993) 1046—1050.

[8] S.E. Esipov, Interface growth and Burgers turbulence: the problem of random initial conditions II. *Phys. Rev.* E 49 (1994) 207—081.

[9] J.B. Grotberg, Pulmonary flow and transport phenomena, *Annu. Rev. Fluid Mech.* 26 (1994) 529—571.

[10] S.N. Gurbatov, A.I. Saichev, Degeneracy of one-dimensional acoustic turbulence for large Reynold numbers, *Sov. Phys.* JETP 80 (1981) 589—595.

[11] S.N. Gurbatov, A.N. Malakhov, A.I. Saichev, *Nonlinear random waves and turbulence in nondispersive media: waves, rays, particles,* Machester University Press, 1991.

[12] S.N. Gurbatov, S.I. Simdyankin, E. Aurell, U. Frisch, G. Toth, On the decay of Burgers turbulence, *J. Fluid Mech.* 344 (1997) 339—374.

[13] Y. Hu, W.A. Woyczynski, Shock density in Burgers' turbulence, In *Nonlinear Stochastic PDEs* (*ed.* T. Funaki and W. A. Woyczynski), 157—165, Springer, 1996.

[14] S. Kida, Asymptotic properties of Burgers turbulence, *J. Fluid Mech.* 93 (1979) 337—377.

[15] M.J. Lighthill, *Viscosity effects in sound waves of finite amplitude*, In Surveys in Mechanics (G.K. Batchelor & R.M. Davids, ed) Cambridge University Press, 250—351, 1956.

[16] G.N. Mercer, A.J. Roberts, A centre manifold description of contaminant dispersion in channels with varying flow properties, *SIAM J. Appl. Math.* 50 (1990) 1547—1565.

[17] G.N. Mercer, A.J. Roberts, A complete model of shear dispersion in pipes, *Jap. J. Indust. Appl. Math.* 11 (1994), 499—521.

[18] S.A. Molchanov, D. Surgailis, W.A. Woyczynski, Hyperbolic asymptotics in Burgers' turbulence and extremal processes, *Commun. Math. Phys.* 168 (1995) 209—226.

[19] J.A. Moriarty, L.W. Schwartz, E.O. Tuck, Unsteady spreading of thin liquid films with small surface tension, *Phys. Fluids* A 3 (1991) 733—742.

[20] J.A. Moriarty, L.W. Schwartz, Dynamic considerations in the closing and opening of holes in thin liquid film, *J. Coil. Interj. Sci.* 161 (1993) 335—342.

[21] A.J. Roberts, Appropriate initial conditions for asymptotic descriptions of the long term evolution of dynamical systems, *J. Austral. Math. Soc.* B 31 (1989) 48—75.

[22] A.J. Roberts, Boundary conditions for approximate differential equations, *J. Austral. Math. Soc.* B 34 (1992) 54—80.

[23] A.J. Roberts, Low-dimensional modelling of dynamics via computer algebra, *Comput. Phys. Comm.* 100 (1997) 215—230.

[24] A.J. Roberts, Low-dimensional models of thin film fluid dynamics, *Phys. Letts.* A 212 (1996) 63—72.

[25] A.J. Roberts, Computer algebra derives correct initial conditions for low-dimensional dynamical models, *Comput. Phys. Comm.* 126 (2000) 215—230.

[26] P.L. Sachdev, K.R.C. Nair, Generalized Burgers equation and Euler-Painleve transcendents II, *J. Math. Phys.* 28 (1987) 997—1004.

[27] P.L. Sachdev, K.R.C. Nair, V.G. Tikekar, Generalized Burgers equation and Euler-Painleve transcendents III, *J. Math. Phys.* 29 (1988) 2397—2400.

[28] J.F. Scott, The long time asymptotics of solution to the generalized Burgers equation, *Proc. R. Soc. Lond.* A 373 (1981) 443—456.

[29] C.G. Wayne, Invariant manifolds and the asymptotics of parbolic equations in cylindrical domains, *Proc. of the first US/China Conference on Differential Equations,* Hangzhou, China, International Press, 1997.

[30] G.B. Whitham, *Linear* and *nonlinear waves,* Wiley, 1974.